\documentclass[11pt]{article}
\usepackage{anyfontsize}
\usepackage{comment}
\usepackage{amsmath}
\DeclareMathOperator{\arcsinh}{arcsinh}
\DeclareMathOperator{\arccosh}{arccosh}
\DeclareMathOperator{\arccoth}{arccoth}
\DeclareMathOperator{\arctanh}{arctanh}

\usepackage[mathscr]{eucal}
\usepackage{amsmath,amsfonts,amssymb,amsthm}
\usepackage[matrix,arrow]{xy}
\usepackage{mathabx}
\usepackage{amsthm,amsmath,latexsym,amssymb,amsfonts,amsbsy}
\usepackage{wrapfig}
\usepackage{graphicx}
\usepackage{float}
\usepackage{soul}

\usepackage{physics}

\usepackage{amsmath}
\usepackage{amssymb}
\usepackage{braket}
\usepackage[T1]{fontenc}
\usepackage{sidecap}

\usepackage{tikz,pgf}
\usetikzlibrary{shapes}
\usetikzlibrary{calc}
\usetikzlibrary{decorations.pathmorphing}
\usetikzlibrary{decorations.pathreplacing,shapes.misc}
\usetikzlibrary{positioning}
\usetikzlibrary{arrows}
\usetikzlibrary{decorations.markings}
\usetikzlibrary{shadings}

\usetikzlibrary{intersections}

\def\be{\begin{equation}}
\def\ee{\end{equation}}
\def\ba{\begin{array}}
\def\ea{\end{array}}

\def\dps{\displaystyle}

\def\arccot{\text{arccot}}

\def\1{\tilde{1}}
\def\2{\tilde{2}}
\def\3{\tilde{3}}

\newdimen\tableauside\tableauside=1.0ex
\newdimen\tableaurule\tableaurule=0.4pt
\newdimen\tableaustep
\def\phantomhrule#1{\hbox{\vbox to0pt{\hrule height\tableaurule
width#1\vss}}}
\def\phantomvrule#1{\vbox{\hbox to0pt{\vrule width\tableaurule
height#1\hss}}}
\def\sqr{\vbox{%
  \phantomhrule\tableaustep

\hbox{\phantomvrule\tableaustep\kern\tableaustep\phantomvrule\tableaustep}%
  \hbox{\vbox{\phantomhrule\tableauside}\kern-\tableaurule}}}
\def\squares#1{\hbox{\count0=#1\noindent\loop\sqr
  \advance\count0 by-1 \ifnum\count0>0\repeat}}
\def\tableau#1{\vcenter{\offinterlineskip
  \tableaustep=\tableauside\advance\tableaustep by-\tableaurule
  \kern\normallineskip\hbox
    {\kern\normallineskip\vbox
      {\gettableau#1 0 }%
     \kern\normallineskip\kern\tableaurule}%
  \kern\normallineskip\kern\tableaurule}}
\def\gettableau#1 {\ifnum#1=0\let\next=\null\else
  \squares{#1}\let\next=\gettableau\fi\next}

\renewcommand{\tilde}{\widetilde}
\renewcommand{\hat}{\widehat}

\def\cO{\mathcal{O}}

\numberwithin{equation}{section} \makeatletter
\@addtoreset{equation}{section}

\def\be{\begin{equation}}
\def\ee{\end{equation}}
\def\ba{\begin{array}}
\def\ea{\end{array}}

\def\dps{\displaystyle}

\def\ba{\begin{array}}
\def\ea{\end{array}}

\def\dps{\displaystyle}

\usepackage{jheppub}

\makeatletter
\def\@fpheader{\vspace{-.1cm}}
\makeatother

\title{Semiclassical torus blocks  in the t-channel}

\author{Juan Ramos Cabezas}

\affiliation{Department of General and Applied Physics, \\
Moscow Institute of Physics and Technology, \\
Institutskiy per. 7, Dolgoprudnyi, \\141700 Moscow region, Russia}

\emailAdd{kabesas.ramos@phystech.edu}

\abstract{We explicitly demonstrate the relation between the 2-point t-channel torus block in the large-$c$ regime and the geodesic length of a specific geodesic diagram stretched in the thermal $AdS_3$ spacetime. 
}

\arxivnumber{}

\begin{document}

\maketitle
\flushbottom

\section{Introduction }

Nowadays the AdS$_3$/CFT$_2$ correspondence is an active field of study. There are many results, particularly in the case of the semi-classical limits where the central charge tends to infinity or, equivalently, the gravitational coupling is small \cite{Brown:1986nw}.  Classical (large-$c$) conformal blocks have been studied in the Riemann sphere, where they were identified with lengths of geodesic networks embedded in the asymptotic AdS spacetime with the angle deficit or BTZ black hole \cite{Hartman:2013mia, Fitzpatrick:2014vua, Asplund:2014coa, Caputa:2014eta, Hijano:2015rla, Fitzpatrick:2015zha, Perlmutter:2015iya, Alkalaev:2015wia, Hijano:2015qja, Alkalaev:2015lca, Beccaria:2015shq, Fitzpatrick:2015dlt, Banerjee:2016qca, Kraus:2016nwo, Chen:2016dfb, Alkalaev:2016rjl,Fitzpatrick:2016ive, Fitzpatrick:2016mtp, Fitzpatrick:2016mjq,Belavin:2017atm, Kusuki:2018J1, Alkalaev:2018nik, Parikh:2019hdfc, Anous:2019pse, Alkalaev:2019zhs, Jepsen:2019svc, Alekseev:2019gkl,  Alkalaev:2020kxz}. Recently, in works \cite{Alkalaev:2016ptm,Alkalaev:2016fok, Kraus:2017ezw, Alkalaev:2017bzx, AlkalaevBelavin:2018J} the holographic dual interpretation of the 1-point and 2- points torus conformal blocks was considered.

It is known that the general solution to the classical euclidean AdS$_3$ gravity is topologically associated with a solid torus (see e.g. \cite{Carlip:1994gc}) so that the corresponding boundary CFT$_2$  lives on a torus \cite{Maloney:2007ud}. In this paper, we take up again the discussion in \cite{Alkalaev:2017bzx, Alkalaev:2016ptm} about the duality of the linearized classical torus conformal block. For the previous studies of the torus conformal blocks in the framework of CFT see \cite{Fateev:2009me, Poghossian:2009mk, Hadasz:2009db, Fateev:2009aw, Menotti:2010en, Marshakov:2010fx, KashaniPoor:2012wb, Piatek:2013ifa}.

\begin{figure}[H]
\centering
\begin{tikzpicture}[scale=0.6][line width=1pt]
\draw (-5,4) node {$\text{Thermal-Euclidean AdS$_3$}$};
\draw[fill=lightgray] (-5,-1.0) circle (0.6cm);
\draw (-5,-1.0) circle (4cm);
\draw[blue] (-5,-0.8) circle (1.4cm);
\draw[blue] (-5,0.6) -- (-5, 1.5);
\fill[blue] (-5,0.6) circle (0.8mm);
\fill[blue] (-5,1.5) circle (0.8mm);
\draw[blue] (-5,1.5) parabola (-3.5, 2.708);
\fill[blue] (-3.5,2.708) circle (0.8mm);
\draw (-3.5,3.2) node {\text{\large y$_2$}};

\draw[blue] (-5,1.5) parabola (-6.5, 2.708);
\fill[blue] (-6.5,2.708) circle (0.8mm);
\draw (-6.5,3.2) node {\text{\large y$_1$}};
\draw (-3.3,-0.8) node {\text{\large 1}};
\draw (-4.7,1) node {\text{\large 2}};
\draw (-4.2,2.2) node {\text{\large 3}};
\draw (-5.7,2.2) node {\text{\large 4}};
\end{tikzpicture}
\caption{The black interior and exterior circles represent the boundary of the thermal AdS. Numbers represent each geodesic trajectory. The first trajectory is a closed trajectory, the second one is a radial trajectory, the third and fourth one are external tajectories attached to the boundary.}
\label{fig:00}
\end{figure}
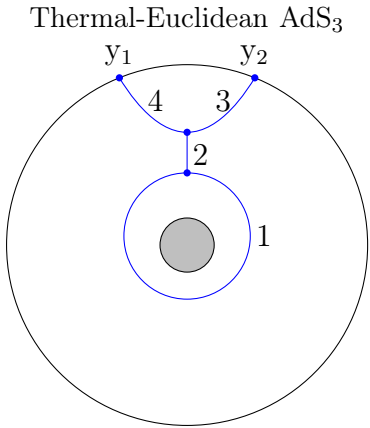

The $c$-large expansions of the conformal blocks are referred to as semiclassical blocks. In \cite{Alkalaev:2016fok} are defined 4 types of semiclassical blocks. One of these semiclassical blocks is the so-called classical block. The original conformal block is the exponential of the classical conformal block in the limits: $c\rightarrow \infty$, and  the relations $\frac{\Delta_i}{c}$ (where $\Delta_i$ stands for all conformal dimensions, on which the conformal blocks depend) are fixed (they are finite different from zero), see \cite{Alkalaev:2016fok}.
The duality proposed in \cite{Alkalaev:2016ptm, Kraus:2017ezw,Alkalaev:2017bzx} describes the linearized classical torus conformal block (we will explain in section \text{\large \ref{sec:cgb}} what the word "linearized" means in this context) as the geodesic length of a specific geodesic diagram embedded in the thermal AdS$_3$ spacetime. This duality translates to the following relation
\begin{equation}
-\text{\Large $g$}^{lin} = S_{thermal} + \sum_{i}\epsilon_iS_i , \label{eq:0}
\end{equation}
where the LHS is the linearized classical torus conformal\footnote{in this paper we use the term linearized large-$\Delta$ instead of linearized classical. This comment will be remarked where we consider that it is necessary} block, while the first term of RHS is the holomorphic part of the 3d gravity action evaluated on the thermal AdS spacetime, and the second term of RHS represents the  sum of the geodesic lengths ($S_i$) over all parts of the diagram (we denote each part of the diagram by index $i$),  each geodesic length ($S_i$) in the sum is multiplied  by its respective mass-parameter ($\epsilon_i$) (these parameters are associated with the conformal dimensions of the block). Fig.\ref{fig:00} represents the diagram, which is associated with the 2-point t-channel torus conformal block. 
In \cite{Alkalaev:2016ptm,Alkalaev:2017bzx} the block/length duality  was demonstrated by explicit  computation in the first order in external conformal dimensions of both sides in (\ref{eq:0}) for the 1-point torus conformal block and 2-point s-channel torus conformal block cases respectively. In \cite{Kraus:2017ezw} the duality was demonstrated for $n$-point global blocks with large dimensions by establishing that the block and length functions satisfy the same equation both on the boundary and in the bulk. In this paper, having in mind these general arguments, we {\it explicitly} calculate and compare the 2-point block function in the t-channel and the length of the corresponding geodesic network in the bulk thereby demonstrating the relation (\ref{eq:0}) explicitly (since the holomorphic part of the  2-point conformal block depends on 4 conformal dimensions, we for simplicity impose that 3 of them are equal to each other, these are the two conformal dimensions of the primary conformal fields and an intermediate conformal dimension. We consider the other intermediate conformal dimension (the loop intermediate conformal dimension) to be much greater than these first three). In particular, we will see that the validity of (\ref{eq:0}) for the 2-point t-channel torus global block case induces the validity of (\ref{eq:0}) for the 1-point torus conformal block case. This is because the linearized classical 1-point torus conformal block is a part of linearized classical 2-point t-channel torus global block. In \cite{Alekseev:2019gkl} the non-perturbative computation of the RHS of (\ref{eq:0}) has been developed for the 1-point torus conformal block case, and we will see that our results agree with those of \cite{Alekseev:2019gkl}.

The order of this paper is the following. In section \text{\large \ref{sec:tcb1}} we review the 2-point torus correlation function and the corresponding 2-point torus conformal blocks, focusing specifically on the t-channel case. In section  \text{\large \ref{sec:gtb1}} we review the computation of the torus global block from the 2-point torus conformal block. In section  \text{\large \ref{sec:cgb}}, introducing the rescaled conformal dimensions we compute the global block with large dimensions which we refer to as  large-$\Delta$ global block. As we have remarked before, our computation is based on the following assumptions: (a) all conformal  dimensions except for the loop intermediate dimension are equal to each other  (b) the loop intermediate dimension is much larger than other dimensions (we detonate the corresponding rescaled loop intermediate dimension by $\tilde{\epsilon}_1$). Thus, expanding the large-$\Delta$ global block up to the $\tilde{\epsilon}_1$-linear term, we obtain the {\it linearized} large-$\Delta$ global block. In section  \text{\large \ref{sec:di1}} we establish the holographic duality for the 2-point linearized large-$\Delta$ global block. We review the prescription of how to calculate the geodesic length in the thermal AdS$_3$ spacetime and characterize the geodesic diagram which length calculates this block. In section \text{\large \ref{sec:glttd}}, we describe in detail the geodesic diagram and find its total geodesic length.  Then, the resulting function is compared with the 2-point linearized large-$\Delta$ global block in t-channel. We close with a brief conclusion in section \text{\large \ref{conclu1}}. All technical details are in appendices \text{\normalsize \ref{sec:app1}}, \text{\normalsize \ref{sec:appendix2}}, \text{\normalsize \ref{app:4}}. 

\section{Torus conformal block}
\label{sec:tcb1}

Let us consider conformal primary fields $\Phi_{\Delta_1, \bar{\Delta}_1}, \Phi_{\Delta_2, \bar{\Delta}_2}$ with conformal weights ($\Delta_1,\bar{\Delta}_1$), ($\Delta_2,\bar{\Delta}_2$). The 2-point correlation function on a torus takes the form 

\begin{equation}
\begin{split}
\braket{\Phi_{\Delta_1, \bar{\Delta}_1}(z_1, \bar{z}_1)\Phi_{\Delta_2, \bar{\Delta}_2}(z_2, \bar{z}_2)} & = Tr(e^{-(Im\tau)\hat{H}+i(Re\tau)\hat{P}}\Phi_{\Delta_1, \bar{\Delta}_1}(z_1, \bar{z}_1)\Phi_{\Delta_2, \bar{\Delta}_2}(z_2, \bar{z}_2))=\\ &= (q\bar{q})^{\frac{-c}{24}}Tr(q^{L_0}\bar{q}^{\bar{L}_0}\Phi_{\Delta_1, \bar{\Delta}_1}(z_1, \bar{z}_1)\Phi_{\Delta_2, \bar{\Delta}_2}(z_2, \bar{z}_2))\label{tcb:1},
\end{split}
\end{equation}
where: $(z_1, \bar{z}_1), (z_2, \bar{z}_2)$ are the coordinates of the conformal primary fields on the plane CFT of the torus, as we will see at the end of this section, there has to be a relation between these coordinates and the coordinates of the AdS$_3$ boundary;  $\hat{H}=2\pi(L_{0}+\bar{L}_{0})-\frac{\pi c}{6}$ is the Hamiltonian of the system; $\hat{P}=2\pi(L_{0}-\bar{L}_{0})$ is
the momentum operator; $q=e^{2\pi i\tau}$, $\tau$ is the torus modular
parameter. The trace in (\ref{tcb:1}) can be calculated in the standard basis of Verma module:
\begin{equation}
|\tilde{\Delta}_1, \bar{\tilde{\Delta}}_1, M,\bar{M}\rangle =\bar{L}^{j_1}_{-n_1}...\bar{L}^{j_l}_{-n_l} L^{i_1}_{-m_1}...L^{i_k}_{-m_{k}}|\tilde{\Delta}_1, \bar{\tilde{\Delta}}_1 \rangle\ , \label{tcb:1-1}
\end{equation}
where $|\tilde{\Delta}, \bar{\tilde{\Delta}}_1\rangle$ is a primary state, $(j_k, n_k, i_k, m_k) \in \mathbb{N}$, ${}\forall k\in \mathbb{N}$ and $M= \sum_k i_km_{k},\, \bar{M}=\sum_k j_kn_k$. The ket  $|\tilde{\Delta}_1, \bar{\tilde{\Delta}}_1, M,\bar{M}\rangle$ is the M-th level descendant state in the Verma module generated from the primary state $|\tilde{\Delta}_1, \bar{\tilde{\Delta}}_1\rangle$. Taking the trace (\ref{tcb:1}) over the states (\ref{tcb:1-1}),  we obtain 

\begin{equation}
\begin{split}
& \braket{\Phi_{\Delta_1, \bar{\Delta}_1}(z_1, \bar{z}_1)\Phi_{\Delta_2, \bar{\Delta}_2}(z_2, \bar{z}_2)} = (q\bar{q})^{\frac{-c}{24}} \sum_{\tilde{\Delta}_1, \bar{\tilde{\Delta}}_1} \sum_{m=0, \bar{m}=0}q^{\tilde{\Delta}_1+m}\bar{q}^{\bar{\tilde{\Delta}}_1+\bar{m}}\times\\ 
& \times \sum_{m=M=N}\sum_{\bar{m}=\bar{M}=\bar{N}}G^{M|N}G^{\bar{M}|\bar{N}}F(\Delta_1, \bar{\Delta}_1, \Delta_2, \bar{\Delta}_2, \tilde{\Delta}_1, \bar{\tilde{\Delta}}_1, M,N, \bar{M}, \bar{N}|z_1,z_2)\label{tcb:2}
\end{split},
\end{equation}
where $G^{M|N}$ is the inverse of the Gram matrix, and the function $F$ is defined as:
\begin{equation}
\begin{split}
& F(\Delta_1, \bar{\Delta}_1, \Delta_2, \bar{\Delta}_2, \tilde{\Delta}_1, \bar{\tilde{\Delta}}_1, M,N, \bar{M}, \bar{N}|z_1,z_2) =\\ &=  \langle \tilde{\Delta}_1, \bar{\tilde{\Delta}}_1, M,\bar{M}|\Phi_{\Delta_1, \bar{\Delta}_1}(z_1, \bar{z}_1)\Phi_{\Delta_2, \bar{\Delta}_2}(z_2, \bar{z}_2)|N,\bar{N}, \tilde{\Delta}_1, \bar{\tilde{\Delta}}_1\rangle.\label{tcb:3}
\end{split}
\end{equation}

\begin{figure}[H]
\centering

\begin{tikzpicture}[scale=0.6, line width=1pt, every node/.style={scale=0.6}]


\draw[black] (-1,0)--(1,0);
\draw (0,0.3) node {$\tilde{\Delta}_2$};

\draw[black] (-2.5,1.5)--(-1,0);
\draw (-1.75,1.3) node {$\Delta_1$};

\draw[black] (-2.5,-1.5)--(-1,0);
\draw (-1.75,-1.3) node {$\Delta_2$};

\fill[black] (-1,0) circle (0.8mm);
\draw (1.75,1.5) node {$\tilde{\Delta}_1$};

\draw[black] (2.5,1.5)--(1,0);
\draw (1.75,-1.3) node {$\tilde{\Delta}_1$};

\draw[black] (2.5,-1.5)--(1,0);
\fill[black] (1,0) circle (0.8mm);
\draw[blue, dashed] (2.5,1.5) .. controls (3.5,0.75) and (3.5,-0.75) .. (2.5,-1.5);

\end{tikzpicture}
\caption{Diagrammatic representation of the t-channel expansion of the 2-point torus conformal block, the two initial primary fields $\Phi_{\Delta_1}, \Phi_{\Delta_2}$ decompose in an intermediate conformal field $\Phi_{\tilde{\Delta}_2}$, and then the latter acts on the descendant states of the intermediate conformal field $\Phi_{\tilde{\Delta}_1}$.}
\label{fig:2}
\end{figure}
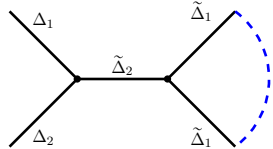

We can expand matrix elements (\ref{tcb:3}) in two different channels, in the so-called s-channel and t-channel. We will focus on the t-channel expansion which  is defined by replacing $\Phi_{\Delta_1, \bar{\Delta}_1}(z_1, \bar{z}_1)\Phi_{\Delta_2, \bar{\Delta}_2}(z_2, \bar{z}_2)$ in (\ref{tcb:3}) with its OPE. Fig.\ref{fig:2} shows the diagrammatic representation of the t-channel expansion of the 2-point conformal block.

The holomorphic part of the OPE of two primary conformal fields is defined as:
\begin{equation}
\Phi_{\Delta_1,\bar{\Delta}_1}(z_1, \bar{z}_1) \Phi_{\Delta_2,\bar{\Delta}_2}(z_2, \bar{z}_2)  = \sum_{\tilde{\Delta}_2}C_{\Delta_1\Delta_2 \tilde{\Delta}_2}(z_1-z_2)^{\tilde{\Delta}_2-\Delta_1-\Delta_2}\psi_{\tilde{\Delta}_2}(z_1,z_2),\label{tcb:4}
\end{equation}
where coefficient $C_{\Delta_1\Delta_2 \tilde{\Delta}_2}$ is the holomorphic part of $ \langle \tilde{\Delta}_2, \bar{\tilde{\Delta}}_2 |\Phi_{\Delta_1,\bar{\Delta}_1}(z_1, \bar{z}_1)|\Delta_2, \bar{\Delta}_2\rangle \times \\\times z_1^{-\tilde{\Delta}_2+\Delta_1+\Delta_2}\bar{z}_1^{-\bar{\tilde{\Delta}}_2+\bar{\Delta}_1+\bar{\Delta}_2}$ and the operator  $\psi_{\tilde{\Delta}_2}(z_1,z_2)$ is 
\begin{equation}\
\psi_{\tilde{\Delta}_2}(z_1,z_2) = \Phi_{\tilde{\Delta}_2}(z_2, \bar{z}_2)+ \beta_1(z_1-z_2)L_{-1}\Phi_{\tilde{\Delta}_2}(z_2, \bar{z}_2)+..., \qquad \beta_1 = \frac{\tilde{\Delta}_2+\Delta_1-\Delta_2}{2\tilde{\Delta}_2}\label{tcb:5}.
\end{equation}
Plugging the OPE (\ref{tcb:4}) into (\ref{tcb:3}) and then placing (\ref{tcb:3}) in (\ref{tcb:2}), we obtain 
\begin{equation}
\begin{split}
  &\langle \Phi_{\Delta_1,\bar{\Delta}_1}(z_1, \bar{z}_1) \Phi_{\Delta_2,\bar{\Delta}_2}(z_2, \bar{z}_2) \rangle   = \sum_{\tilde{\Delta}_1, \bar{\tilde{\Delta}}_1}\sum_{\tilde{\Delta}_2, \bar{\tilde{\Delta}}_2}C_{\Delta_1\Delta_2\tilde{\Delta}_2}C_{\bar{\Delta}_1\bar{\Delta}_2\bar{\tilde{\Delta}}_2} \times \\
  & \times z_{1}^{-\Delta_1}\bar{z}_1^{-\bar{\Delta}_1} z_{2}^{-\Delta_2}\bar{z}_2^{-\bar{\Delta}_2}[\, \nu_c^{\Delta_{1,2}, \tilde{\Delta}_{1,2}}(q,z_1,z_2) \nu_c^{\bar{\Delta}_{1,2}, \bar{\tilde{\Delta}}_{1,2}}(\bar{q},\bar{z}_1,\bar{z}_2)]\,,
\end{split} \label{tcb:6}
\end{equation}
where $\tilde{\Delta}_1, \tilde{\Delta}_2$ are intermediate  conformal dimensions and the t-channel (holomorphic) conformal block is 
\begin{equation}
\begin{split}
\nu_c^{\Delta_{1,2}, \tilde{\Delta}_{1,2}}(q,z_1,z_2) & = q^{-c/24+\tilde{\Delta}_1}(\frac{z_1}{z_2})^{\Delta_1}(\frac{z_1-z_2}{z_2})^{\tilde{\Delta}_2-\Delta_1-\Delta_2} \sum_{n=0}^{\infty}q^n \times\\ & \times \sum_{|M|=|N|=n}G^{M|N}\frac{\langle\tilde{\Delta}_1, M|\psi_{\tilde{\Delta}_2}(z_1, z_2)|N,\tilde{\Delta}_1\rangle}{\langle\tilde{\Delta}_1|\Phi_{\tilde{\Delta}_2}(z_2)|\tilde{\Delta}_1\rangle}. \label{tcb:7}
\end{split}
\end{equation}
In the context of the block/length duality which we want to demonstrate, the coordinates $z= (z_1,z_2)$ of the conformal block (\ref{tcb:7}) are given by a conformal map $z(X)$ from the AdS$_3$ boundary to the plane CFT, here $X$ are the coordinates at the AdS$_3$ boundary.

\section{Torus global blocks}
\label{sec:gtb1}

The global  t-channel torus block is given by (\ref{tcb:7}), where we consider only  the generators $L_{-1}, L_1, L_0$ (and their antiholomorphic counterparts), which are the generators of $SL(2, \mathbb{C})$ group. In \cite{Alkalaev:2017bzx} the general formula for the global t-channel torus block is given . 
The 2-point t-channel global block is
\begin{equation}
\begin{split}
V^{\Delta_{1,2}, \tilde{\Delta}_{1,2}}(q, w) &= q^{-c/24+\tilde{\Delta}_1}(1-w)^{\Delta_1}(-w)^{-\Delta_1-\Delta_2-\tilde{\Delta}_2} \times \\ &\times \sum_{n,m=0}^{\infty}\frac{(\tilde{\Delta}_2+\Delta_1-\Delta_2)_{(m)}(\tilde{\Delta}_2)_{(m)}\tau_{n,n}(\tilde{\Delta}_1, \tilde{\Delta}_2, \tilde{\Delta}_1)q^n (-w)^{m}}{n!m!(2\tilde{\Delta}_1)_n (2\tilde{\Delta}_2)_m},\label{gtb:1}
\end{split}
\end{equation}
where
\begin{equation}
 w= \frac{z_2-z_1}{z_2}, \label{gtbextra1:1}
\end{equation}
$(a)_{(n)}$ is the Pochhammer symbol for a number $a \in \mathbb{C}$, and \cite{Alkalaev:2015fbw}
\begin{equation}
\begin{split}
\tau_{n,m}(x_1,x_2,x_3) = \sum_{p=0}^{min[n,m]}C_p^{n}(2x_3+m-1)^{(p)}(m)^{(p)} & (x_3+x_2-x_1)_{(m-p)}\times \\&\times(x_1+x_2-x_3+p-m)_{(n-p)},
\end{split}
\end{equation}
$(a)^{(n)} = C_n^{a}n!$ and $C_k^{n}$ is the combinatorial coefficient. Hereafter we omit the factor $q^{-c/24}$.

Using the  hypergeometric functions we can express (\ref{gtb:1}) in a concise form  \cite{Kraus:2017ezw},
\begin{equation}
\begin{split}
V^{\Delta_{1,2}, \tilde{\Delta}_{1,2}}(q, w)= & \frac{q^{\tilde{\Delta}_1}}{(1-q)^{1-\tilde{\Delta}_2}} {}_2F_1(\tilde{\Delta}_2, \tilde{\Delta}_2+2\tilde{\Delta}_1-1, 2\tilde{\Delta}_1,q)\times \\ & \times w^{-\Delta_1-\Delta_2+\tilde{\Delta}_2}(1-w)^{\Delta_1}{}_2F_1(\tilde{\Delta}_2, \tilde{\Delta}_2-\Delta_1+\Delta_2, 2\tilde{\Delta}_2,w) \label{gtb:2},
\end{split}
\end{equation}
We must note that the first line of (\ref{gtb:2}) is exactly the 1-point global global block given in  \cite{Hadasz:2009db,Alkalaev:2016fok}. 

\section{Large-$\Delta$ global blocks}
\label{sec:cgb}
We begin this section by making some observations regarding equation (\ref{gtb:2}), and some previous results of the conformal blocks. In works \cite{Alkalaev:2016ptm, Alkalaev:2017bzx} was shown perturbatively that the linearized classical version of the 1-point torus conformal block (linearized classical 1-point torus conformal block) was associated with geodesic length of a certain diagram embedded in the euclidean thermal AdS$_3$ spacetime. On the other hand, in work \cite{Alkalaev:2016fok} the relation that exists between the linearized classical (in our terms linearized large-$\Delta$) version of the 1-point torus global block with the linearized classical 1-point torus conformal block was conjectured,  that relation is given by equation (5.9) of work \cite{Alkalaev:2016fok}, that equation defines a direct relation between the linearized classical (linearized large-$\Delta$) 1-point torus global block and the linearized classical 1-point torus conformal block. In this sense, the holographic correspondence of the linearized classical 1-point torus conformal block of works \cite{Alkalaev:2016ptm, Alkalaev:2017bzx} is at the same time a correspondence for the linearized classical (linearized large-$\Delta$) 1-point torus global block.

The main objective of this work is to find a holographic correspondence of the global block (\ref{gtb:2}),  more precisely, to find a correspondence for a certain version of global block (\ref{gtb:2}). Since block (\ref{gtb:2}) is factorized in terms of 1-point torus global block,  and based on the above mentioned with respect to the linearized classical (linearized large-$\Delta$) 1-point torus global block, we consider that it is necessary to look for this correspondence in the large-$\Delta$ version of the global block (\ref{gtb:2}) (more precisely in the linearized large-$\Delta$ version). In this sense, we will find the linearized large-$\Delta$ version of block (\ref{gtb:2}).

By the definition of the 2-point large-$\Delta$ global block we have \cite{Alkalaev:2017bzx,Alkalaev:2016fok}
\begin{equation}
V^{\Delta_{1,2}, \tilde{\Delta}_{1,2}}(q, w) = exp[k\hspace{0.1cm} \text{\Large $g$}_{2pt}^{la}(\epsilon_i, \tilde{\epsilon}_j,q,w)], \quad \quad as \quad k\rightarrow \infty, \label{cgb:1}
\end{equation}
where $\text{\Large $g$}_{2pt}^{la}(\epsilon_i, \tilde{\epsilon}_i,q,w)$ is the large-$\Delta$ global block, and 
\begin{equation}
\Delta_i = k \epsilon_i, \qquad \tilde{\Delta}_i = k \tilde{\epsilon}_i, 
\end{equation}
where $\epsilon_i, \tilde{\epsilon}_i$ are rescaled conformal dimensions (In \cite{Alkalaev:2016fok} they are called classical global dimensions ), in our case $i = 1,2$, and $k$ as we defined it in (\ref{cgb:1}) is a large dimensionless parameter.

In what follows we consider one of the most simple cases when the conformal dimensions are constrained as
\begin{equation}
\label{dim}
\epsilon_1= \epsilon_2= \tilde{\epsilon}_2 = \epsilon, \qquad \frac{\epsilon}{\tilde{\epsilon}_1}= \delta << 1. 
\end{equation}
We want to analyze the behavior of $\text{\Large $g$}_{2pt}^{la}$ under these conditions and then obtain the so-called linearized large-$\Delta$ global block which we denote by  $\text{\Large $g$}_{2pt}^{lin}$. In this paper we define the {\it linearized} large-$\Delta$ global block as the $\tilde{\epsilon}_1$-linear part of the large-$\Delta$ global block $\text{\Large $g$}^{la}_{2pt}$, this definition translates to: 
\begin{equation}
\text{\Large $g$}_{2pt}^{la}(\epsilon_i, \tilde{\epsilon}_i,q,w) = \text{\Large $g$}_{2pt}^{lin}(\tilde{\epsilon}_1,\delta,q,,w) + \cO(\tilde{\epsilon}_1^2).\label{cgb:2}
\end{equation}
Note that the definition of the linearized block depends on the largest dimension which we select, since the linearized block will be the part (of the block) which is linear in the largest dimension. 

Calculating $\text{\Large $g$}_{2pt}^{la}$ using the definitions (\ref{gtb:2}) and  (\ref{cgb:1}) and taking its $\tilde{\epsilon}_1$-linear part, we obtain that
\begin{equation}
\text{\Large $g$}_{2pt}^{lin}(\tilde{\epsilon}_1,\delta,q,,w) = \text{\Large $g$}_{1pt}^{lin}(\tilde{\epsilon}_1,\delta,q)+ \delta\tilde{\epsilon}_1(\log{\frac{1-w}{w}}-2(\arccosh{\frac{1}{\sqrt{w}}}-\log{\frac{2}{\sqrt{w}}})),\label{cgb:3}
\end{equation}
where $\text{\Large $g$}_{1pt}^{lin}(\tilde{\epsilon}_1,\delta,q)$ is the 1-point linearized large-$\Delta$ global block given in \cite{Alkalaev:2016ptm, Alkalaev:2016fok},(see appendix \ref{sec:app1}). In \cite{Alkalaev:2016fok} $\text{\Large $g$}_{1pt}^{lin}(\tilde{\epsilon}_1,\delta,q)$ was obtained by taking the large-$k$ asymptotic expansion of the 1-point global block, and it was conjectured that $\text{\Large $g$}_{1pt}^{lin}(\tilde{\epsilon}_1,\delta,q)$ is equal to the linearized classical 1-point conformal block of \cite{Alkalaev:2016ptm}. Coming back to complex variables $z_1$ and $z_2$ in (\ref{cgb:3}) through (\ref{gtbextra1:1}), and representing  $z_1=re^{iy_1}, z_2=re^{iy_2}$ we obtain the final expression
\begin{equation}
\text{\Large $g$}_{2pt}^{lin}(\tilde{\epsilon}_1,\delta,q,y_1,y_2) = \text{\Large $g$}_{1pt}^{lin}(\tilde{\epsilon}_1,\delta,q)- \delta\tilde{\epsilon}_1\log[{(\cosh{i\frac{y_1-y_2}{4}})^2\sinh{i\frac{y_1-y_2}{2}}}]\;.\label{cgb:4}
\end{equation}

\section{Dual interpretation}
\label{sec:di1}

In analogy to works \cite{Alkalaev:2016ptm,Kraus:2017ezw,Alkalaev:2017bzx}, the holographic dual interpretation of the linearized large-$\Delta$ global block is  that the 2-point linearized large-$\Delta$ global block (\ref{cgb:4}) is associated with  the geodesic length of a diagram (trajectories) embedded in the $AdS_3$ spacetime and  shown in Fig.\ref{fig:1}. We call this diagram the t-channel diagram for the 2-point conformal block. It consists  of four  parts: a loop part, a radial part, and two external parts. There is another diagram associated with the 2-point conformal block, it is the s-channel diagram treated in \cite{Alkalaev:2017bzx}. All these diagrams are embedded in the Euclidean thermal $AdS_3$ spacetime 
\begin{equation}
    ds^2 = (1+\frac{r^2}{l^2})dt^2+(1+\frac{r^2}{l^2})^{-1}dr^2+r^2d\varphi^2, \label{hdi:1}
\end{equation}
where $t\sim  t + \beta, \varphi \sim \varphi + 2\pi, r \geq 0$. We set $l=1$. The time period $\beta$ is associated with the modular parameter $\tau_{ads}=i\beta/2\pi$ and the temperature $\beta\sim T^{-1}$.

In the geodesic approximation, the gravity functional integral is calculated near the
saddle-point given by a particular solution. At low temperatures, or in other words for Im$(\tau_{ads}) \gg 1$, the thermal AdS predominates in the functional integral \cite{Maldacena:1998bw}, and therefore, in this case, the classical action is determined mainly by the gravitational part. It has been calculated in \cite{Carlip:1994gc, Maldacena:1998bw,Kraus:2006wn} to be $S_{thermal}= \frac{i\pi\tau_{ads}}{2}$. Furthermore, there is a matter part of the action, which we associate with geodesic lengths of worldlines of  massive particles (the masses are the rescaled conformal dimensions $\tilde{\epsilon}_{1,2}, \epsilon_{1,2})$. As a result, the total on-shell classical action is given by the gravitational and matter parts as 
\begin{equation}
    S_{total}=S_{thermal}+\tilde{\epsilon}_1S_{loop}+\tilde{\epsilon}_2 S_{radial}+\epsilon S_{external}. \label{hdi:2}
\end{equation}
Here, the first term is the gravitational part of the action in the thermal AdS, the second, third and fourth terms correspond to the material part of the action, they  are the geodesic lengths of the loop trajectory, radial line and external trajectories respectively, see Fig. \ref{fig:1}. 


\subsection{Worldline approach}

The worldline approach has been effective for the calculation of lengths of various geodesic networks \cite{Hijano:2015rla, Alkalaev:2015wia, Banerjee:2016qca}. Each geodesic segment is given by the following action:
\begin{equation}
    S= \int_{1}^{2}d\lambda\sqrt{g_{mn}\dot{x}^m\dot{x}^n},\label{worldap:1}
\end{equation}
where 1 and 2 are initial/final positions, local coordinates are $x^m= (t,\varphi,r)$, the metric coefficients $g_{mn}(x)$ we can obtain from (\ref{hdi:1}), and the velocities $\dot{x}^m = (\frac{dt}{d\lambda}, \frac{d \varphi}{d\lambda}, \frac{d r}{d\lambda}) = (\dot{t},\dot{\varphi}, \dot{r} )$ are defined with respect to the evolution parameter $\lambda$, their covariant versions are the corresponding momenta, which are defined as:

\begin{equation}
p_{t} = g_{tt}\dot{t}, \quad  p_{\varphi} = g_{\varphi \varphi}\dot{\varphi}, \quad p_{r} = g_{rr}\dot{r}.
\label{worldap:1-1}
\end{equation}
The action is reparametrization invariant, therefore we can set the following normalization condition:
\begin{equation}
|g_{mn}\dot{x}^m\dot{x}^n|= 1,\label{worldap:2}
\end{equation}
so that from (\ref{worldap:1}) we obtain the on-shell value of the action  $S= \lambda_2-\lambda_1$.

Since the metric coefficients do not depend on time and angular variables the respective momenta are constant $\dot{p}_t=0$,  $\dot{p}_{\varphi}=0$. We restrict the dynamics to the level surface (as a condition) characterized by the constant angle $\varphi=0$ corresponding to conserved $p_{\varphi}=0$ while the other conserved momentum $p_t$ is the motion integral giving the shape of a geodesics curve. From the normalization condition (\ref{worldap:2}) we have $g_{tt}\dot{t}^2+g_{rr}\dot{r}^2=1$ and taking into account $g_{tt}g_{rr}=1$, we have:
\begin{equation}
\dot{r}= \pm\sqrt{r^2-s^2+1}, \qquad    s= |p_t|, \label{worldap:3}
\end{equation}
the overall sign $\pm$ tells us if $r$ decreases (sign $-$) or increases (sing $+$) along the proper time $\lambda$. The length  of the loop trajectory can be calculated using the definition of the time momentum $p_t = g_{tt}\dot{t}$, from which we have
\begin{equation}
\dot{t}= \frac{s}{1+r^2},  \qquad S_{loop} = \frac{1}{s}\int_0^{\beta}dt(1+r^2(t)), \label{worldap:4}
\end{equation}
where $s$ is the loop momentum, and $r(t)$ is the radial deviation. From equation (\ref{worldap:3}) and (\ref{worldap:4}) we can calculate the derivative of $r$ with respect to time $t$,  $\frac{dr}{dt} = \frac{dr}{d\lambda}\frac{d\lambda}{dt}$, so we have the following equation of evolution:
\begin{equation}
\frac{dr}{dt}= \pm\frac{1}{s}(1+r^2)\sqrt{r^2-s^2+1}. \label{worldap:5}
\end{equation}
Here, the signs $\pm$ are from (\ref{worldap:3}), the selected sign says us if the function $r(t)$ increases or decreases. We can find also the length of trajectory contained between two points $r_1$ and $r_2$ through the equation
\begin{equation}
S= \int_{r_1}^{r_2}\frac{dr}{\sqrt{1+r^2-s^2}} = \log{\sqrt{1+r_2^2-s^2}}-\log{\sqrt{1+r_1^2-s^2}},\label{world:6}
\end{equation}
\subsection{Equilibrium condition on vertex points}
In each vertex point (an intersection of three geodesic lines) there is   a sum of three terms of the type (\ref{worldap:1}). Minimizing the combination of these terms we find out that the time momenta satisfy the following weighted equilibrium condition (see \cite{Alkalaev:2017bzx} for more details):
\begin{equation}
\tilde{\epsilon}p_m^1 + \tilde{\epsilon}p_m^2 + \epsilon p_m^0=0, \label{world:7}
\end{equation}
where $p^{1,2}$ are the ingoing/outgoing momenta of geodesic segments with the same mass parameter and $p^0$ is an external momentum (geodesic segments with a different mass parameter in relation to the previous two), and $m={(t,r)}$.

\section {Geodesic length and t-channel global block} \label{sec:glttd}

In the t-channel case, the whole diagram of trajectories is composed of four parts, each part is interpreted as the path of a classical particle  that moves in the thermal Euclidean AdS$_3$ spacetime with a mass equal to rescaled conformal dimension ($\epsilon$ or $\tilde{\epsilon}$).  The first part is the loop trajectory beginning from point ($t_0,\rho_1$) and going  around $t$ (in the clockwise direction), finally ending at the same point ($t_0,\rho_1$). In Fig. \ref{fig:1} we denote the radius of this trajectory, its momentum and its mass-parameter by $R, s$  and $\tilde{\epsilon}_1$ respectively.  The second part is the radial trajectory that is connected to the loop trajectory at the point $(t_0,\rho_1)$ and from this point goes radially to the point ($t_0,\rho_2$), its momentum and mass-parameter are denoted by $s_0$ and $\tilde{\epsilon}_2$. The third and fourth parts are two external trajectories,  both of them start at the point ($t_0,\rho_2$), the fourth one ends at point ($y_2, R_2=\infty$) and the third one ends at the point ($y_1, R_1=\infty$) ($y_2<y_1$). In the general case their masses can be different, but we will just analyze the case when these  two parameters are the same. We denote this mass  by $\epsilon$, in the Fig. \ref{fig:1}, and  their radius and momenta by $R_1, R_2, s_1, s_2$, respectively. Quantities $R(t), R_1(t), R_2(t),t_0, s,s_0,s_1,s_2,\rho_1,\rho_2$ are functions of variables $\tilde{\epsilon}_1, \tilde{\epsilon}_2,\epsilon, y_1, y_2$. 

%
\subsection{System of equations}
\label{sysequation1:1}
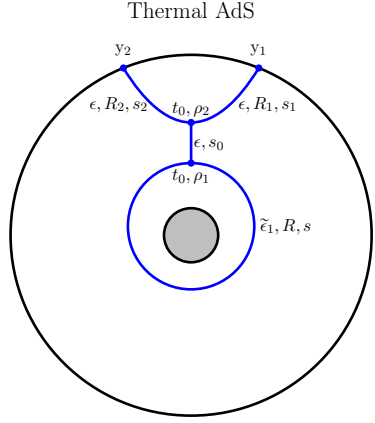
\begin{figure}[H]
\centering

\begin{tikzpicture}[scale=0.6, line width=1pt, every node/.style={scale=0.6}]

\draw (-5,4) node {$\text {\Large Thermal AdS}$};
\draw[fill=lightgray] (-5,-1.0) circle (0.6cm);
\draw (-5,-1.0) circle (4cm);
\draw[blue] (-5,-0.8) circle (1.4cm);
\draw[blue] (-5,0.6) -- (-5, 1.5);
\fill[blue] (-5,0.6) circle (0.8mm);
\draw (-5,0.3) node {$t_0, \rho_1$};
\fill[blue] (-5,1.5) circle (0.8mm);
\draw (-5,1.8) node {$t_0, \rho_2$};
\draw[blue] (-5,1.5) parabola (-3.5, 2.708);
\fill[blue] (-3.5,2.708) circle (0.8mm);
\draw (-3.5,3.1) node {$\text{y$_1$}$};
\draw[blue] (-5,1.5) parabola (-6.5, 2.708);
\fill[blue] (-6.5,2.708) circle (0.8mm);
\draw (-6.5,3.1) node {$\text{y$_2$}$};
\draw (-2.9,-0.8) node {$\tilde{\epsilon}_1,R, s$};
\draw (-4.6, 1) node {$\epsilon,  s_0$};
\draw (-3.3,1.9) node {$\epsilon, R_1,s_1$};
\draw (-6.6,1.9) node {$\epsilon, R_2,s_2$};

\end{tikzpicture}
\caption{The black interior and exterior circles represent the boundary of the thermal AdS. Points (t$_0$,$\rho_1$) and  (t$_0$,$\rho_2$) are two vertices, $\epsilon$ and $\tilde{\epsilon}_1$ denote  masses  of each segment, functions  $R,R_1, R_2$ denote the radii of the geodesic lines, $s, s_0, s_1, s_2$ denote angular momenta.}
\label{fig:1}
\end{figure}

In this section we basically follow \cite{Alkalaev:2017bzx}. Fig. \ref{fig:1} and the respective system of equations describe the t-channel geodesic diagram for the 2-point global block.

The equation of motion for the radius is given by
 \begin{equation}
\dot{R} = \frac{1+R^2}{s}\sqrt{R^2-s^2+1} \label{borrar:0}.
\end{equation}
Integrating (\ref{borrar:0}) for the loop trajectory and taking into account the initial condition (\ref{borrar:2}) we find  
 \begin{equation}
\begin{split}
& e^{2(t-t_0)} = \\&=\frac{(i+\rho_1)(R(t)-i)(\rho_1-s\sqrt{\rho_1^2-s^2+1}+is^2-i)(s\sqrt{R(t)^2-s^2+1}+R(t)-is^2+i)}{(i-\rho_1)(R(t)+i)(\rho_1+s\sqrt{\rho_1^2-s^2+1}-is^2+i)(s\sqrt{R(t)^2-s^2+1}-R(t)-is^2+i)} \label{borrar:1},
\end{split}
\end{equation}
\begin{equation}
R(\beta) = R(t_0) = \rho_1. \label{borrar:2}
\end{equation}

The momenta  equilibrium equations at the vertex point $(t_0,\rho_1)$ are
\begin{equation}
\tilde{\epsilon}_1(s-s) + \tilde{\epsilon}_2s_0=0, \qquad 2\tilde{\epsilon}_1\sqrt{\rho_1^2-s^2+1}-\tilde{\epsilon}_2\sqrt{\rho_1^2-s_0^2+1}=0. \label{borrar:3}
\end{equation}
For the external trajectories we fix the initial point of $R_1(t)$ and $R_2(t)$ as $t=t_0$, while  $R_1(y_1)=\infty$, $R_2(y_2)=\infty$ along with $R_1(t_0)= R_2(t_0)=\rho_2$ are boundary conditions. Integrating (\ref{borrar:0}) for radial functions $R_1(t), R_2(t)$ and imposing the above mentioned conditions, we arrive at 
\begin{equation}
 e^{2(y_1-t_0)} = \frac{\rho_2^2+s_1^2(\rho_2^2-1)-2\rho_2s_1\sqrt{\rho_2^2-s_1^2+1}}{(\rho_2^2+1)(s_1-1)^2},\label{borrar:4}
\end{equation}
\begin{equation}
 e^{2(y_2-t_0)} = \frac{\rho_2^2+s_2^2(\rho_2^2-1)-2\rho_2s_2\sqrt{\rho_2^2-s_2^2+1}}{(\rho_2^2+1)(s_2-1)^2}.\label{borrar:5}
\end{equation}

The momentum equilibrium equations at the point $(t_0, \rho_2)$ are 
\begin{equation}
\epsilon s_1+\epsilon s_2+\epsilon s_0=0, \qquad \epsilon\sqrt{\rho_2^2-s_1^2+1}+ \epsilon\sqrt{\rho_2^2-s_2^2+1} -\epsilon\sqrt{\rho_2^2+1}=0,\label{borrar:6}
\end{equation}
where we used \eqref{dim}.

In order to calculate geodesic length we need explicit expressions for  $R(t),s,\rho_1,s_0,t_0, s_1,s_2,\rho_2$ that solve  the defining system (\ref{borrar:1})-(\ref{borrar:6}). In the next sections we consider  these equations in more detail and give solutions. 

\subsection{Solving the defining system} 
\label{subsec:ltr1}
Firstly, we are going to compute the lengths of the loop and radial trajectories. Since these lengths do not depend on $t_0$, we can set $t_0 = 0$ for simplicity so that (\ref{borrar:2}) becomes
\begin{equation}
R(\beta) = R(t_0=0) = \rho_1. \label{gltt:3}
\end{equation}
From (\ref{borrar:3}) we have $s_0=0$ and
 \begin{equation}
\rho_1 = \sqrt{\frac{s^2}{1-\delta^2/4}-1}, \qquad where \qquad \delta = \epsilon/\tilde{\epsilon}_1\label{gltt:5}. 
 \end{equation}
 We have three equations (\ref{borrar:1}), (\ref{gltt:3}), (\ref{gltt:5}) to find three unknown functions $R(t),\rho_1, s$. We will refer to this system of equations as 1-point system of equations because it describes the geodesic diagram of  1-point conformal block \cite{Alkalaev:2016ptm}. There are two independent solution to this system of equations, both yielding  the same geodesic length, see Appendix \ref{sec:appendix2} for more details. Here we present only one solution, 
 \begin{equation}
 \begin{split}
& s = \sqrt{\frac{\delta^2}{4\sin^2{(\beta/2)}}+1}, \qquad \rho_1 = \frac{\delta\coth{\beta/2}}{\sqrt{4-\delta^2}}, \\
& R(t) =  \frac{-\delta(\sinh(t-\beta)-\sinh{(t)})}{\sqrt{(1-\cosh{(\beta)})(4-\delta^2+ \delta^2\cosh{(2t-\beta)}-4\cosh{(\beta}))}}\label{gltt:6}.
\end{split}
 \end{equation}
 Finding the geodesic length of this part of the diagram, we obtain 
\begin{equation}
L= \tilde{\epsilon}_1(L_{loop}+\delta L_{radial}),
\label{gltt:6.1}
\end{equation} 
\begin{equation}
\dps
L_{loop} = \frac{1}{s}\int_{0}^{\beta}(R(t)^2+1)dt,
\label{gltt:6.2}
\end{equation} 
\begin{equation}
L_{radial}= \int_{\rho_1}^{\rho_2}\frac{1}{\sqrt{r^2+1}}dr = \arcsinh[\rho_2]-\arcsinh[\rho_1],
\label{gltt:6.3}
\end{equation}
and, finally, 
 \begin{equation}
 \begin{split}
L =\frac{\log{q}}{4}+ \tilde{\epsilon}_1[2\arccoth{\frac{1+q}{\sqrt{-1+2q-q^2-q\delta^2}}} & +  \delta\log{\frac{\delta(1+q) + 2\sqrt{1-2q+q^2+q\delta^2}}{(-1+q)\sqrt{4-\delta^2}}}]  +\\ &+\tilde{\epsilon}_1\delta \arcsinh[\rho_2]\;. \label{gltt:7} 
\end{split}
 \end{equation}
For further convenience, we separate the 1-point part of this expression as 
 \begin{equation}
\begin{split}
 L_{1pt-diagram} = \frac{\log{q}}{4} &+ \tilde{\epsilon}_1(2\arccoth{\frac{1+q}{\sqrt{-1+2q-q^2-q\delta^2}}}  +\\ &+  \delta\log{\frac{\delta(1+q) + 2\sqrt{1-2q+q^2+q\delta^2}}{(-1+q)\sqrt{4-\delta^2}}}).
 \label{gltt:7-2} 
\end{split}
 \end{equation}
 In the limit $\rho_2\rightarrow\infty$ in \eqref{gltt:7} the first line of this equation correspond to the  geodesic length of the 1-point diagram, this result coincides with those given in  \cite{Alkalaev:2016ptm,Alekseev:2019gkl}. 

Now, we are going to compute the geodesic lengths of the external trajectories. We know that $s_0=0$, so that using  this relation  in (\ref{borrar:6}) we obtain  $s_1=-s_2$. Thus, taking the ratio  (\ref{borrar:4})$/$(\ref{borrar:5}) and using $\rho_2(s_1)$ found  from (\ref{borrar:6}), finally, we obtain an equation for $s_1$. Solving this equation we find $s_1$ and then $\rho_2$ as 
\begin{equation}
\rho_2= \frac{2\csch{\frac{y_1-y_2}{2}}-\coth{\frac{y_1-y_2}{2}}}{\sqrt{3}},\qquad s_1=s_2= \coth{\frac{y_1-y_2}{2}}- \frac{1}{2}\csch{\frac{y_1-y_2}{2}}.\label{gltt:10-2}
\end{equation}
The lengths  of external trajectories are the same due to the reflection symmetry of the geodesic diagram (for equal external dimensions only). It is given by (up to an infinite constant)  
\begin{equation}
L_{external}= 2\epsilon\int_{\rho_3}^{\infty}\frac{1}{\sqrt{R_1^2-s_1^2+1}}dR_1= -2\epsilon\log[\rho_2+ \sqrt{\rho_2^2-s_1^2+1}]. \label{gltt:10-3}
\end{equation}
\subsection{Relation between geodesic length and global block} \label{subsec:tglw}
Using (\ref{gltt:10-2}), (\ref{gltt:10-3}), (\ref{gltt:7}) and (\ref{gltt:7-2}) we  calculate the total geodesic length of the entire diagram which is given by
\begin{equation}
L_{total}= L_{1pt-diagram}+\tilde{\epsilon}_2\arcsinh{\rho_2}+L_{external}, \label{gltt:14.1} 
\end{equation}
where
\begin{equation}
\begin{split}
L_{total} =  \frac{\log{q}}{4}+ \tilde{\epsilon}_1[2\arccoth{\frac{1+q}{\sqrt{-1+2q-q^2-q\delta^2}}} + \delta\log{\frac{\delta(1+q) + 2\sqrt{1-2q+q^2+q\delta^2}}{(-1+q)\sqrt{4-\delta^2}}}]  +\\
+ \tilde{\epsilon}_1\delta(const+ \log{[(\cosh{\frac{y_1-y_2}{4}})^2\sinh{\frac{y_1-y_2}{2}}]}).\label{gltt:15}
\end{split}
\end{equation}
We see the following relation between (\ref{gltt:15}) and (\ref{cgb:4}) up to a constant. 
\begin{equation}
S_{thermal}+L_{total} = -\text{\Large $g$}_{2pt}^{lin}(\tilde{\epsilon}_1,\delta,q,y_1,y_2)|_{y_i\rightarrow -iy_i}. \label{comp:1}
\end{equation}
The equation (\ref{comp:1}) is our main result.

\section{Conclusions} 
\label{conclu1}

In this work, the torus block/length duality proposed in \cite{Alkalaev:2016ptm,Kraus:2017ezw,Alkalaev:2017bzx}  has been confirmed in the 2-point case by finding explicit expressions for the torus block with large dimensions  in t-channel and the corresponding length of dual geodesic diagram. This duality was demonstrated non-perturbatively in conformal dimensions. More precisely, it was possible to verify   the duality relation (\ref{eq:0}) for the linearized large-$\Delta$ 2-point torus global block in the t-channel, the demonstration of this case also implied non-perturbative check of (\ref{eq:0}) for  1-point block. It follows that these blocks have a dual interpretation in terms of certain geodesic diagrams in the thermal AdS$_3$ spacetime. More specifically, the linearized large-$\Delta$ 2-point torus global block is calculated by the geodesic length of the diagram shown in Fig. \ref{fig:1} by means of the equation (\ref{comp:1}).

To conclude, let us shortly mention that explicit  calculations  in the $s$-channel case  entails a greater degree of complexity. This is  because the respective algebraic equation  system of section \text{\large \ref{sysequation1:1}} is of degrees greater than four.

\hfill\break
\textbf{Acknowledgments.} The author thanks the following people very deeply: K.B. Alkalaev for having proposed the problem discussed here and for his constant ideas that helped to improve this work; M.M. Pavlov for useful discussions; his family for their unconditional support.

\appendix
\section{Linearized large-$\Delta$ block }
\label{sec:app1}
In this appendix  we discuss equation (\ref{cgb:3}) in more detail. From (\ref{cgb:1}) it follows that
\begin{equation}
\text{\Large $g$}_{2pt}^{la}(\tilde{\epsilon}_{1,2}, \epsilon_{1,2}, q, w) = \lim_{k\rightarrow\infty} \frac{\log{V^{\Delta_{1,2}\tilde{\Delta}_{1,2}}(q,w)}}{k}\bigg|_{
(\tilde{\Delta}_{1,2},\Delta_{1,2})  \rightarrow ( k\tilde{\epsilon}_{1,2}, k\epsilon_{1,2})}. \label{A:1}
\end{equation}
Taking into account (\ref{gtb:2}) we find  that
\begin{equation}
\text{\Large $g$}_{2pt}^{la}(\tilde{\epsilon}_{1,2}, \epsilon_{1,2},q,w) = \text{\Large $g$}_{1pt}^{la}(\tilde{\epsilon}_1,\delta,q) +\epsilon\log{\frac{1-w}{w}}+ \lim_{k\rightarrow\infty}\frac{\log{{}_2F_1(k\tilde{\epsilon}_1\delta, k\tilde{\epsilon}_1\delta, 2k\tilde{\epsilon}_1\delta,w)}}{k},
\label{A:2}
\end{equation}
where $\epsilon_1= \epsilon_2= \tilde{\epsilon}_2 =\epsilon$, $\delta =\frac{\epsilon}{\tilde{\epsilon}_1}<<1$, and $\text{\Large $g$}_{1pt}^{la}(\tilde{\epsilon}_1,\delta,q)$ is the 1-point large-$\Delta$ global block 
\begin{equation}
\text{\Large $g$}_{1pt}^{la}(\tilde{\epsilon}_1,\delta,q)=\lim_{k\rightarrow\infty}\frac{\frac{q^{k\tilde{\epsilon}_1}}{(1-q)^{1-k\tilde{\epsilon}_2}}{}_2F_1(k\tilde{\epsilon}_1\delta, k\tilde{\epsilon}_1\delta+2k\tilde{\epsilon}_1-1,2k\tilde{\epsilon}_1, q)}{k}. \label{A:3}
\end{equation}
The 1-point large-$\Delta$ global block will be studied in appendix \ref{app:4}, here we will refer only to the second and third terms of (\ref{A:2}). Taking the $\tilde{\epsilon}_1$-linear part of (\ref{A:2}) we get the linearized large-$\Delta$ 2-point global block $\text{\Large $g$}_{2pt}^{lin}$. To do this, we expand the third term of (\ref{A:2}) in terms of $w$, take the limit $k\rightarrow \infty$, and then keep just $\tilde{\epsilon}_1$-linear terms. Finally, we obtain that 
\begin{equation}
\text{\Large $g$}_{2pt}^{lin}(\tilde{\epsilon}_{1,2}, \epsilon_{1,2},q,w) = g_{1pt}^{lin}(\tilde{\epsilon}_1,\delta,q) + \tilde{\epsilon}_1\delta\log{\frac{1-w}{w}}+ \tilde{\epsilon}_1\delta\sum_{n=1}\frac{(2n-1)!!w^n}{(2n)!!!n}, \label{A:4}
\end{equation}
where $(2n-1)!!= (2n-1)(2n-3)(2n-5)...(1)$, and $(2n)!!!= (2n)(2n-2)(2n-4)...(2)$. Rewriting (\ref{A:4}) in terms of $\arccosh{(w})$ we obtain (\ref{cgb:3}) 
\begin{equation}
\text{\Large $g$}_{2pt}^{lin}(\tilde{\epsilon}_{1,2}, \epsilon_{1,2},q,w) = \text{\Large $g$}_{1pt}^{lin}(\tilde{\epsilon}_1,\delta,q)+ \delta\tilde{\epsilon}_1(\log{\frac{1-w}{w}}-2(\arccosh{\frac{1}{\sqrt{w}}}-\log{\frac{2}{\sqrt{w}}})), \label{A:5}
\end{equation}
where $\text{\Large $g$}_{1pt}^{lin}(\tilde{\epsilon}_1,\delta,q)$ is the $\tilde{\epsilon}_1$-linear part of (\ref{A:3}), which we will discuss in appendix \ref{app:4} in more detail.

\section{Geodesic length of 1-point conformal diagram } 
\label{sec:appendix2}

\subsection{Exact solutions for the 1-point diagram system of equations}

In this section, we will solve the following system of equations 
\begin{equation}
\begin{cases}
 (\ref{borrar:1}), \\ (\ref{gltt:3}), \\ (\ref{gltt:5}),
 \end{cases} \label{exactsol:1}
\end{equation}
for $R(t)$, $\rho_1$ and $s$ in terms of $t$, $\delta$, $\beta$. This system of equations is solved in \cite{Alkalaev:2016ptm, Alkalaev:2017bzx} perturbatively.
Making the following change of variables 
\begin{equation}
A= \frac{(-1+i\rho_1)(-i\rho_1+is\sqrt{\rho_1^2-s^2+1}+s^2-1)e^{-2t}}{(\rho_1-i)(\rho_1+s\sqrt{\rho_1^2-s^2+1}-is^2+i)}, \label{exactsol:2}
\end{equation}
\hfill \break
the equation (\ref{borrar:1}) can be rewritten as an equation of fourth degree in terms of $R(t)$ (remember that in section (\ref{subsec:ltr1} ) we set $t_0=0$ 
\begin{equation}
\begin{split}
R^4(t)(-1+2A-A^2+s^2+2As^2+A^2s^2)&+ 2R^2(t)(-1+2A-A^2+s^2+A^2s^2)+\\&+(-1+2A-A^2+s^2-2As^2+A^2s^2)=0.\label{exactsol:3}
\end{split}
\end{equation}
Solutions of (\ref{exactsol:3}) are
\begin{equation}
R_{1,2}(t)= \pm\frac{(-1+A)\sqrt{1-s^2}}{\sqrt{-1+2A-A^2+s^2-2As^2+A^2s^2}}, \qquad R_{3,4} = \pm i. \label{exactsol:4}
\end{equation}
Solution $R_{3,4}$ are not physical because they are pure complex. Using (\ref{exactsol:4}) we rewrite (\ref{gltt:3}), obtaining a new equation in terms of $\rho_1$, and $s$, then replacing $\rho_1$ by (\ref{gltt:5}) in this new equation we obtain an equation just in terms of $s$, solving this equation we obtain $s$. There are 3 different solution of $s$
\begin{equation}
s^{(1)} = \sqrt{1 + \frac{\delta^2 }{4\sin^2(\beta/2)}}, \qquad  s^{(2)} = \sqrt{1 - \frac{\delta^2}{4\cos^2(\beta/2)}}, \qquad s^{(3)} = 0. \label{exactsol:5}
\end{equation}
If we substitute $s^{(3)}$ into (\ref{gltt:5}) we will have $\rho_1= i$, therefore, this solution is not physical. We find two different solutions for $\rho_1$
\begin{equation}
\rho_1^{(1)} = \frac{\delta\coth(\beta/2)}{\sqrt{4-\delta^2}},  \qquad
\rho_1^{(2)} = \frac{\delta\tanh(\beta/2)}{\sqrt{4-\delta^2}}. \label{exactsol:6}
\end{equation}
Using (\ref{exactsol:4}) for each pair of ($\rho_1,s$) we find a general solution of $R(t)$
\begin{equation}
\begin{split}
& R_1(t) = \frac{\delta(\sinh(t-\beta)-\sinh{(t)})}{\sqrt{(1-\cosh{(\beta)})(4-\delta^2+ \delta^2\cosh{(2t-\beta)}-4\cosh{(\beta}))}}.\\
&  R_2(t)= \frac{\sqrt{2} \delta \sinh(|t-\beta/2|)}{\sqrt{4-\delta^2-\delta^2\cosh(2t-\beta)+4\cosh\beta}}.\label{exactsol:7}
\end{split}
\end{equation}
\newline
\hfill \break
Summarizing, we have:

First solution
\begin{equation}
\begin{split}
& s_1 = \sqrt{1 + \frac{\delta^2 }{4\sin^2(\beta/2)}},\qquad \rho_1^{(1)} = \frac{\delta \coth{\beta/2}}{\sqrt{4-\delta^2}}, \\ 
& R_1(t) = \frac{\delta(\sinh(t-\beta)-\sinh{(t)})}{\sqrt{(1-\cosh{(\beta)})(4-\delta^2+ \delta^2\cosh{(2t-\beta)}-4\cosh{(\beta}))}}. \label{exactsol:8}
\end{split}
\end{equation}

Second solution
\begin{equation}
\begin{split}
& s_2 = \sqrt{1 - \frac{\delta^2}{4\cos^2(\beta/2)}}, \qquad \rho_1^{(2)} = \frac{\delta \tanh{\beta/2}}{\sqrt{4-\delta^2}},\\
& R_2(t)= \frac{\sqrt{2} \delta \sinh(|t-\beta/2|)}{\sqrt{4-\delta^2-\delta^2\cosh(2t-\beta)+4\cosh\beta}}. \label{exactsol:9}
\end{split}
\end{equation}

\subsection{Geodesic lengths}

In this section, we  compute the geodesic lengths of the loop-trajectory and radial trajectory of the diagram of Fig.\ref{fig:1}. The part containing $\rho_2$ in (\ref{gltt:6.3}) will not be  considered. We calculate the geodesic length of the loop-trajectory and radial trajectory with the equations (\ref{gltt:6.2}) and (\ref{gltt:6.3}), respectively. 

For the First solution we have
\begin{equation}
\begin{split}
&S_{loop} =\frac{\arctanh(\frac{2\cosh(\beta/2)}{\sqrt{-2+\delta^2+2\cosh(\beta)}})\sqrt{-2+\delta^2+2\cosh(\beta)}\csch(\beta/2)}{\sqrt{1+\frac{\delta^2\csch^2(\beta/2)}{4}}}, \\ & S_{radial} =-\arcsinh(\frac{\delta \coth(\beta/2)}{\sqrt{4-\delta^2}}), \label{exactsol:10}
\end{split}
\end{equation}
\begin{equation}
\begin{split}
 &S_{1pt-diagram} =\\ 
 &  \begin{split}=  & -\beta/4 + \tilde{\epsilon}(\frac{\arctanh(\frac{2\cosh(\beta/2)}{\sqrt{-2+\delta^2+2\cosh(\beta)}})\sqrt{-2+\delta^2+2\cosh(\beta)}\csch(\beta/2)}{\sqrt{1+\frac{\delta^2\csch^2(\beta/2)}{4}}} -\\ &- \delta \arcsinh(\frac{\delta \coth(\beta/2)}{\sqrt{4-\delta^2}})).\label{exactsol:11}
\end{split}
\end{split}
\end{equation}
Let us express  $S_{1pt-diagram}$ in terms of $q= e^{2i\pi\tau}$ ($\tau$ is the CFT modular parameter of the torus ) provided that $\beta = -\log{q_{ads}}$ and $q_{ads} = e^{2\pi i\tau_{ads}}$. For this solution we set $\tau_{ads} = \tau$,\footnote{as explained in \cite{Alkalaev:2016ptm,Alkalaev:2017bzx}} which means that $q= q_{ads}$ and $\beta = -\log{q}$. Then, we obtain
\begin{equation}
\begin{split}
S_{1pt-diagram}  =\frac{\log{q}}{4}&+ \tilde{\epsilon}_1[2\arccot{\frac{1+q}{\sqrt{-1+2q-q^2-q\delta^2}}} +\\&+ \delta\log{\frac{\delta(1+q) + 2\sqrt{1-2q+q^2+q\delta^2}}{(-1+q)\sqrt{4-\delta^2}}}].\label{exactsol:12}
\end{split}
\end{equation}

For the Second solution, we have
\begin{equation}
\begin{split}
& S_{loop} =  -\frac{\arctanh(\frac{(1+\cosh(\beta))\tanh(\beta/2)}{\sqrt{\cosh^2(\beta/2)(2-\delta^2+2\cosh(\beta))}})(-2+\delta^2-2\cosh(\beta))}{\sqrt{\cosh^2(\beta/2)(2-\delta^2+2\cosh(\beta))(1-(\delta \sech{(\beta/2)})^2/4)}},\\ & S_{radial} =-\arcsinh(\frac{\delta\tanh(\beta/2)}{\sqrt{4-\delta^2}}), \label{exactsol:13}
\end{split}
\end{equation}
\begin{equation}
\begin{split}
& S_{1pt-diagram} =\\ & \begin{split}= -\beta/4 & +  \tilde{\epsilon}_1[-\frac{\arctanh(\frac{(1+\cosh(\beta))\tanh(\beta/2)}{\sqrt{\cosh^2(\beta/2)(2-\delta^2+2\cosh(\beta))}})(-2+\delta^2-2\cosh(\beta))}{\sqrt{\cosh^2(\beta/2)(2-\delta^2+2\cosh(\beta))(1-(\delta \sech{(\beta/2)})^2/4)}}-\\ &-\delta \arcsinh(\frac{\delta\tanh(\beta/2)}{\sqrt{4-\delta^2}}) ]. \label{exactsol:14}
\end{split}
\end{split}
\end{equation}
For this solution we set $\tau = \tau_{ads}+1/2$, this means that $\beta = -\log{q_{ads}}= -\log{(-q)}$. Making this change of variables for $\tau_{ads}$ we obtain the following equation for $S_{1pt-diagram}$
\begin{equation}
\begin{split}
S_{1pt-diagram}  =\frac{\log{q}}{4}&+ \tilde{\epsilon}_1[-2\arctan{\frac{1+q}{\sqrt{-1+2q-q^2-q\delta^2}}} +\\&+ \delta\log{\frac{\delta(1+q) + 2\sqrt{1-2q+q^2+q\delta^2}}{(-1+q)\sqrt{4-\delta^2}}}].\label{exactsol:15}
\end{split}
\end{equation}

$S_{1pt-diagram}$ from (\ref{exactsol:12}) and (\ref{exactsol:15}) are equal since the expansion in Taylor series of $\arctan{x}$ and $-\arccot{x}$  coincide in all orders of $\delta^n$ for $n\geq1$. The zeroth term ($\delta^0$) for both (\ref{exactsol:12}) and (\ref{exactsol:15}) is $-(\tilde{\epsilon}_1-1/4)\log{q}$.

\section{1-point global case}
\label{app:4}
In this section we  demonstrate that the linearized large-$\Delta$ 1-point global block  is equal to the geodesic length of the 1-point diagram (\ref{gltt:7-2}). The 1-point global block was analyzed in \cite{Alkalaev:2017bzx, Alkalaev:2016fok,Kraus:2017ezw, Hadasz:2009db}, its general formula was given in \cite{Alkalaev:2017bzx, Alkalaev:2016fok, Kraus:2017ezw} in different forms, the following is one of them 
 \begin{equation}
V_{1pt}(\tilde{\Delta}_1, \tilde{\Delta}_2,q) = \frac{q^{\tilde{\Delta}_1}}{(1-q)^{1-\tilde{\Delta}_2}} {}_2F_1(\tilde{\Delta}_2, \tilde{\Delta}_2+2\tilde{\Delta}_1-1, 2\tilde{\Delta}_1,q).  \label{D:1}
 \end{equation}
where $\tilde{\Delta}_1, \tilde{\Delta}_2$ are intermediate conformal dimensions and their corresponding rescaled conformal dimensions ($\tilde{\epsilon}_1, \tilde{\epsilon}_2$)  correspond to the mass-parameter of the loop and radial trajectories respectively. We want to calculate the large-$\Delta$ global block $\text{\Large $g$}_{1pt}^{la}$ through the definition (\ref{cgb:1})
 \begin{equation}
\text{\Large $g$}_{1pt}^{la}(\tilde{\epsilon}_1, \delta,q)= \lim_{k \rightarrow \infty}\frac{\log{[V_{1pt}(\tilde{\Delta}_1, \tilde{\Delta}_2,q)]}\bigg{|}_{(\tilde{\Delta}_1, \tilde{\Delta}_2) \rightarrow  (k\tilde{\epsilon}_1, k \tilde{\epsilon}_2)}}{k},  \label{D:1-2} 
 \end{equation}
 where,  as before $\tilde{\epsilon}_{1,2}$ are rescaled conformal dimensions, $\delta = \tilde{\epsilon}_2/\tilde{\epsilon}_1$. After the computation of the large-$\Delta$ 1-point global block, we compute the linearized large-$\Delta$ global block $\text{\Large $g$}^{lin}_{1pt}$, which is the $\tilde{\epsilon}_1$-linear part of (\ref{D:1-2}), all this process of the computation of the linearized large-$\Delta$ global block from the global block is explained in \cite{Alkalaev:2016fok}, in fact, the formula (2.12) of \cite{Alkalaev:2016fok} gives a novel representation of the linearized large-$\Delta$ global block, it is called the integral representation of the linearized large-$\Delta$ global block. It has the form\footnote{the term $-\frac{\log{q}}{4}$ of equation (\ref{D:2}) comes from the factor $q^{-c/24}$, and in the limit $k \rightarrow \infty$  we have taken $c/k \rightarrow 6$.} 
\begin{equation}
\text{\Large $g$}_{1pt}^{lin}(\tilde{\epsilon}_1,\delta,q) = (\tilde{\epsilon}_1-\frac{1}{4})\log{q}+ \tilde{\epsilon}_1\int_{0}^{q}dx(-\frac{1}{x}+ \sqrt{\frac{1}{x^2}+\frac{\delta^2}{x(1-x)^2}}). \label{D:2}
\end{equation}
In general, we want to demonstrate that 
\begin{equation}
    (\ref{gltt:7-2})= -(\ref{D:2})+\tilde{\epsilon}_1f(\delta),  \label{D:3}
\end{equation}
where $f(\delta)$ is a function  of $\delta$ only. That can be demonstrated by expanding (\ref{gltt:7-2}) and (\ref{D:2}) in powers  of $\delta$. Expanding ($\ref{D:2}$) we obtain 
\begin{equation}
 \text{\Large $g$}_{1pt}^{lin}(\tilde{\epsilon}_1,\tilde{\epsilon}_2,q) = (\tilde{\epsilon}_1-\frac{1}{4})\log{q} + \tilde{\epsilon}_1 \sum_{n=1}^{\infty}\frac{q^n{}_2F_1(n,2n,1+n,q)}{n}C_{1/2}^{n}\delta^{2n}, \label{D:4}
\end{equation}
where $C_{1/2}^{n}$ is the binomial coefficient $\begin{pmatrix}
1/2\\n
\end{pmatrix} = \frac{(1/2)(1/2-1)..(1/2-n-1)}{n!} $, and ${}_2F_1(n,2n,1+n,q)$ is the ordinary hypergeometric function. Expanding (\ref{gltt:7-2}) we obtain
\begin{equation}
\begin{split}
& L_{1pt-diagram} = \frac{\log{q}}{4}+\\&+ \tilde{\epsilon}_1\sum_{n=0}^{\infty}\frac{C_{-1/2}^{n}(-1+q)^{-2n}q^n(1+q){}_2F_1(1,-n,\frac{1}{2}-n, -\frac{(-1+q)^2)}{4q})\delta^{2(n+1)}}{4(1+n)(1+2n)(q-1)}.  \label{D:5}
\end{split}
\end{equation}
$C_{-1/2}^{n}$ in the binomial coefficient $\begin{pmatrix}
-1/2\\n
\end{pmatrix}$. Finally, we obtain the relation 
\begin{equation}
\begin{split}
-g^{lin}_{1pt}(\tilde{\epsilon}_1,\delta,q)-L_{1pt-diagram}& = \tilde{\epsilon}_1\sum_{n =1}^{\infty}\frac{2^{-2-2(-1+n)}\delta^{(2n)}}{n(1+2(-1+n))} =\\&= \tilde{\epsilon}_1(\delta\arctanh{\frac{\delta}{2}}+\log{\frac{4-\delta^2}{4}}), \label{D:6}
\end{split}
\end{equation}
which demonstrates (\ref{D:3}) and simultaneously fixes $f(\delta)$.
\bibliography{refs}
\end{document}